\documentstyle[12pt]{article}

\def\be{\begin{equation}}
\def\ee{\end{equation}}
\def\ba{\begin{array}{c}}
\def\ea{\end{array}}

\def\ben{$$}
\def\een{$$}
\begin{document}

\titlepage

 \begin{center}{\Large \bf
Spiked potentials and quantum toboggans

 }\end{center}

\vspace{5mm}

 \begin{center}
Miloslav Znojil

\vspace{3mm}

\'{U}stav jadern\'e fyziky\footnote{e-mail: znojil@ujf.cas.cz }
 AV \v{C}R, 250 68 \v{R}e\v{z}, Czech
Republic

\end{center}

\vspace{5mm}

\section*{Abstract}

Even if the motion of a quantum (quasi-)particle proceeds along a
left-right-symmetric (${\cal PT}-$symmetric) curved path ${\cal C}
\neq I\!\!R$ in complex plane $C\!\!\!\!C$, the spectrum of bound
states may remain physical (i.e., real and bounded below). A
generalization is outlined. Firstly, we show how the topologically
less trivial (tobogganic) contours ${\cal C}$ may be allowed to
live on several sheets of a Riemann surface. Secondly, the
specification of a scattering regime is formulated for such a
class of models.

 \vspace{9mm}

\noindent
 PACS 03.65.Ge

%

\newpage

\section{Introduction \label{I} }

Let us be interested in the analytic power-law potentials
 \be
 V(x)=\sum_{\beta}\,g_{(\beta)}\,x^\beta, \ \ \ \ \ \ \
 \beta \in I\!\!R\,,
 \label{nespiked}
 \ee
useful as a schematic phenomenological model as well as a
laboratory for testing numerical methods. Among the most popular
versions of this choice one finds the polynomially perturbed
harmonic oscillator (with even integers $\beta_j=2j$, $j = 1, 2,
\ldots, j_{max}$ and with a comparatively easy perturbative
tractability \cite{Flueggea}) as well as its ``spiked" versions
(admitting negative integer exponents $\beta_k=-k$, $k = 1, 2,
\ldots, k_{max}$ -- cf. \cite{Harrell,Hall}). From this background
one may derive many mathematically equivalent descendants
(\ref{nespiked}) with rational exponents $\beta \in O\!\!\!\!Q$ by
an elementary change of variables $x$ and $\psi(x)$ in the
associated Schr\"{o}dinger equation \cite{classif}.

The existence of the latter equivalence transformations (which, in
essence, has been revealed by Liouville \cite{Liouville}) will
simplify some purely technical aspects of our forthcoming
considerations. In particular, it will enable us to restrict our
attention just to the most common asymptotically harmonic spiked
oscillators
 \be
 V(x)=\frac{g_{(-2)}}{x^2}+\sum_{\beta>-2}\,g_{(\beta)}\,x^\beta + x^2
  ,  \ \ \ \ \ \ \ g_{(-2)}=\ell(\ell +1) > 0
  \,.
 \label{spiked}
 \ee
This sum will be assumed finite and containing just subdominant
anharmonic terms with rational exponents, $\beta \in (-2,2)$.

Our specific choice of the quadratic spike in (\ref{spiked}) is
inspired by its key role in the above-mentioned Liouvillean
changes of variables (cf. also section \ref{IV.1} below) and,
first of all, in their implementation by  Buslaev and Grecchi
\cite{BG}. In the latter paper a set of {\em manifestly
non-Hermitian} Hamiltonians with real spectra has been considered,
with interactions of the form (\ref{nespiked}). Surprisingly
enough, the presence of the spike proved crucial in the rigorous
demonstration of the reality of the spectrum (cf. ref. \cite{BG}
for more details). Another argument supporting the inclusion of a
nontrivial centrifugal-like spike may be found in the more recent
papers by Dorey, Dunning and Tateo \cite{DDT,DDTb} who found a
rigorous proof of the reality of the spectrum which has been
extended to potentials (\ref{spiked}) with $g_{(-2)} \neq 0$ in
very natural manner.

In a marginal remark Buslaev and Grecchi noticed that their
Hamiltonians are ${\cal PT}-$symmetric,
 \be
 H_{({\cal PT})}=p^2+V_{({\cal PT})}(x)
 \neq H_{({\cal PT})}^\dagger \equiv {\cal P}\, H_{({\cal PT})}
 \,{\cal P}^{-1}\,,
 \ee
with parity ${\cal P}$ and complex conjugation  ${\cal T}$ (which
mimics time reversal). In a historical perspective it is a
paradox, therefore, that the present enormous growth of popularity
of ${\cal PT}-$symmetric potentials $V_{({\cal PT})}(x)$ has not
been initiated by the paper \cite{BG} but, a few years later, by
the doubts-breaking letter of Bender and Boettcher \cite{BB} who
emphasized that the analyticity of $V_{({\cal PT})}(x)$ may
represent one of hidden reasons for the surprising reality of the
spectrum \cite{BBjmp}.

During the subsequent quick development of the field it became
clear that one is allowed to define and integrate Schr\"{o}dinger
equation with an analytic potential along the whole
``analytic-continuation" families of complex contours ${\cal C}
\neq I\!\!R$ in complex plane $C\!\!\!\!C$. The ``allowed
deformations" of each particular choice of ${\cal C}$ must not
cross the natural boundaries of the domain of analyticity of the
underlying ``physical" bound-state wavefunction $\psi(x)$
\cite{BB,Tanaka}.

{\it Vice versa}, the change of the boundary conditions may be
expected to imply a nontrivial change of the spectrum. This has
already been noticed, in 1993, by Bender and Turbiner \cite{BT}.
An explicit quantitative numerical verification of the latter
expectation in an elementary ${\cal PT}-$symmetric model with
exponential asymptotics of $V(x)$ may be found in our early study
\cite{Guard}.

In the most popular example $V(x) \sim x^{2+\epsilon}$ of a ${\cal
PT}-$symmetric potential with power-law asymptotics it is
particularly easy to specify the contours ${\cal C} = \lim_{a \to
\infty} \Gamma_a$ as the limits of the finite,
non-selfintersecting and sufficiently smooth contours
$\Gamma_a=\{\zeta (t),|\,t\in (-a,a),a>0\}$ in $C\!\!\!\!C$ (cf.,
e.g., \cite{Tanaka} for all details -- one must set $\lim_{t \to
\pm \infty}=\pm \infty$ etc). What has to be emphasized in
parallel is that even for the above elementary
$\epsilon-$dependent power-law asymptotics of potentials the
requirement of the left-right symmetry still leaves the shape of
the curve $\zeta(t)$ quite ambiguous. Thus, one considers just the
classes of curves $\zeta(t)$ which remain compatible with the {\em
same} asymptotic boundary conditions,
 \be
 {\rm arg}\ x \equiv
 {\rm arg}\ \zeta (t) \in
 \left \{
 \begin{array}{cc}
 \left (
 -\frac{\pi}{4+\epsilon}+\frac{\epsilon\pi}{8+2\pi}-\pi,
 \frac{\pi}{4+\epsilon}+\frac{\epsilon\pi}{8+2\pi}-\pi
 \right ),& t \ll -1,\\
 \left (
 -\frac{\pi}{4+\epsilon}-\frac{\epsilon\pi}{8+2\pi},
 +\frac{\pi}{4+\epsilon}-\frac{\epsilon\pi}{8+2\pi}
 \right ),& t \gg 1
 \ea
 \right .
 \label{wedge}
 \ee
for the large arguments $|t|$ and coordinates $|\zeta(t)|$ and for
not too large $\epsilon$ at least.

In this context our present main message may be formulated as a
generalization of the usual definition of the non-intersecting
${\cal PT}-$symmetric paths ${\cal C} = {\cal C}^{(0)}$ which are
reflection symmetric with respect to the imaginary axis. Our
generalization will be based on the observation that even in the
simplest examples of the type $V(x) \sim x^{2+\epsilon}$ the
``standard" branch point in infinity becomes complemented by
another branch point in the origin (at all irrational or
non-integer $\epsilon>0$). As a consequence, also the wavefunction
$\psi(x) \equiv \psi[\zeta(t)]$ acquires a branch point at $x=0$.
Even if $\epsilon$ remains integer this branch point becomes
generated by the spike in $V(x)$ with $g_{(-2)}\neq 0$.

In the similar situations people usually use an additional
requirement that the whole curve $\zeta(t)$ {\em does not cross
the cut} (chosen, usually, from $x=0$ upwards). In our recent
letter \cite{tobog} we noticed that one could easily remove the
latter restriction and work with the reflection symmetric paths
${\cal C}^{(N)}$ with $N > 0$ (apparent) self-intersections
whenever the wavefunction $\psi(x)$ possesses a branch point
$x^{(BP)}_{0}$ at $x=0$ and ``lives" on several Riemann sheets. In
such a case (cf. the illustrative sample of ${\cal C}^{(2)}$ in
our present Figure 1) the emergence of a nontrivial geometric
phase in $\psi(x)$ after rotation $x \to x \cdot e^{2\pi {\rm i}}$
gives rise to a new, topological source of the possible
modification of the energy spectrum.

In our present paper, the latter possibility is further being
developed. Although our illustrative examples still remain
comparatively simple, we shall, in general, admit the existence of
$x^{(BP)}_{0}$ at $x=0$ as well as of some other $2M$ non-zero
branch points forming the left-right symmetric pairs
$x^{(BP)}_{\pm m}$ with $m = 1, 2, \ldots M$. Their presence will
be again interpreted as reflecting the presence of singularities
of the ${\cal PT}-$symmetric potentials $V(x)$ in complex plane.

In the two brief introductory paragraphs \ref{II.1} and \ref{II.2}
of section \ref{II} we shall assume that
$\max(\beta)=\beta_{max}<2$ and $\min(\beta)=\beta_{min}>-2$ in
eq.~(\ref{spiked}). Such a requirement (closely related, by the
way, to the perturbation-series convergence \cite{Kato}) will help
us to minimize inessential technicalities. As we already noticed,
such an assumption of the relative boundedness of the whole
anharmonic perturbation in (\ref{spiked}) is artificial and can be
relaxed via a change of variables. Nevertheless, its use will help
us to define ${\cal PT}-$symmetry of $\psi(x)$ even in the
presence of branch points.

In section \ref{II.3}, first of all, the emerging ambiguity of the
very concepts of parity ${\cal P}$ and of complex conjugation
${\cal T}$ will be discussed and the method of its elimination
will be clarified. In subsection \ref{IV.1} details will be added
for the first nontrivial case with $M=0$. The emergence of
difficulties at the higher $M> 0$ will then be pointed out in
subsection \ref{IV.2}.

In the subsequent section \ref{III} we shall discuss the
possibilities of the existence of a nontrivial, ${\cal
PT}-$symmetric version of the analytic ${\cal S}-$matrix (cf.
subsection \ref{III.1}). Using a solvable example in subsection
\ref{III.2} we shall finally show that and how this could lead to
a new understanding of the ${\cal PT}-$symmetric systems along
directions where the first steps have already been made by Ahmed
et al \cite{Berry} and by Cannata et al \cite{Berryb,Berryc}.

In section  \ref{IV} we shall summarize the appeal and compactness
of the picture allowing the use of trajectories living on several
Riemann sheets. The persistence of many technical as well as
physical open questions will be emphasized.

\section{Potentials possessing a single singularity
\label{II}}

The presence of the centrifugal term in $V(x) \sim x^{2+\epsilon}$
forms a very natural starting point for the use of transfer
matrices connecting independent-solution pairs over different
asymptotic ``Stokes sectors"  \cite{DDT} or ``wedges"  \cite{BB}
defined by formulae of the type (\ref{wedge}). In this perspective
one may interconnect the techniques of (exact) WKB approximants
\cite{Voros} and of the Bethe ansatz \cite{DDTb} with the language
of monodromy group \cite{Arnold} or of the so called Stokes
geometry \cite{Trinh,proc} as well as with the analyses of
solvable models in quantum field theory \cite{Tateo}.

The first steps are to be made here in the direction where the
eligible Stokes sectors lie on several Riemann sheets pertaining
to the analytic solutions $\psi(x)$ of the differential
Schr\"{o}dinger equations possessing $2M+1$ or $2M$ branch points
$x^{(BP)}_{\pm m} = {\cal PT} x^{(BP)}_{\mp m}{\cal PT}\in
C\!\!\!\!C $ with $m \leq M$ and $m \geq 0$ or $m \geq 1$,
respectively.

\subsection{Solvable example: Tobogganic harmonic oscillator
 \label{II.1}}

The three-dimensional harmonic-oscillator Schr\"{o}dinger equation
with its ordinary differential radial re-scaled form
 \be
 \left (
 -\frac{d^2}{dx^2} + \frac{\ell(\ell+1)}{x^2} + x^2
 \right ) \,\psi (x)= E \,\psi (x)
   \label{HO}
 \ee
is one of the most popular illustrations of the formalism of the
textbook Quantum Mechanics (TQM, \cite{Messiah}). Traditionally,
its bound states are sought in the usual Hilbert space
$L_2(I\!\!R^3)$ and one may make use of the proportionality of the
wave functions to the Laguerre polynomials,
 \be
 \psi(x) = \psi_{n,\ell}^{(TQM)}(x)=
 {\cal N}_{n,\ell}\,
 \,{x^{\ell\,+1}}\,e^{-x^2/2}\,
 L_n^{(\ell\,+1/2)}(x^2),
 \ee
 \ben
 \ \ \ \ \ \ \
 E=E_{n,\ell}^{(TQM)}=4n+2\ell+3, \ \ \ \ \
  \ \ \ \ \ \ \
 n, \ell = 0, 1, \ldots\,.
 \een
Incidentally,  the {\em same} differential eq. (\ref{HO}) may play
the role of an illustrative example in the various consistent
formulations of the contemporary ${\cal PT}-$symmetric Quantum
Mechanics \cite{koneclit}. In eq. (\ref{HO}) one only has to
replace the standard textbook {\em real} domain ${\cal D}^{(TQM)}=
(0,\infty) \equiv I\!\!R^+$ of $x$, say, by the Buslaev's and
Grecchi's \cite{BG} straight contour
 \be
 {\cal C}^{(0)} = {\cal D}_{\varepsilon}^{(PTSQM)}=\left \{
 x\, \left | \, x = t-i\,\varepsilon ,\,
 t\in I\!\!R, \ \varepsilon > 0 \right .
 \right \}
 \label{BGC}
 \ee
which is {\em complex}, left-right symmetric and ``twice as long".
As a consequence, there emerge ``twice as many" bound-state levels
with a not too dissimilar structure. At an arbitrary real
$\alpha(\ell)\equiv \ell +1/2$ and discrete $n = 0, 1, \ldots$ the
new solutions retain the closed and compact form
 \be
 \psi(x) = \psi_{n,\ell,\,\pm}^{(PTSQM)}(x)=
 {\cal N}_{n,\ell,\,\pm}\,
 \,\sqrt{x^{1\pm 2\alpha(\ell)}}\,e^{-x^2/2}\,
 L_n^{[\pm \alpha(\ell)]}(x^2),
 \ee
 \ben
 \ \ \ \ \ \ \
 E=E_{n,\ell, \,\pm}^{(PTSQM)}=4n+2 \pm 2\alpha(\ell)
 \een
and contain now a new discrete quantum number $q=\pm$ of
quasi-parity \cite{ptho}.

\subsection{Perturbed tobogganic harmonic oscillators
 \label{II.2}}

In ref. \cite{tobog} we followed the inspiration by ref. \cite{BG}
and admitted a ``realistic" perturbation $\lambda\,W(ix)$ of the
potential in (\ref{HO}),
 \be
 \left [
 -\frac{d^2}{dx^2} + \frac{\ell(\ell+1)}{x^2} + x^2
  +\lambda\,W(ix)
 \right ] \,\psi (x)= E \,\psi (x)\,.
   \label{AHO}
 \ee
For technical reasons we constrained  our attention to a few
particular $W(ix) \sim \sum_\beta\,g_\beta \,(ix)^\beta$. On this
background we imagined that after the perturbation the {\em same }
differential equation (\ref{AHO}) may generate {\em different}
spectra on different contours.

Basically, what we did was that from the presence of the
centrifugal-like singularity in the potential we deduced the
topological nontriviality of the Riemann surface on which our wave
function $\psi(x)$ was defined by the differential eq.
(\ref{AHO}). In what follows we intend to deduce a few further
consequences of that observation.

{\it A priori}, one might link this type of analysis to monodromy
properties (of geometric phase type) of solutions at the branch
point. This appealing possibility (mentioned also by the anonymous
referee of this paper) is already under an active consideration in
the project \cite{myinprep}. In the simpler approach the idea will
be pursued here via a replacement of the Buslaev's and Grecchi's
straight contour (\ref{BGC}) (and/or of all its admissible
analytic-continuation deformations ${\cal C}^{(BG)} $) by the much
broader family of the topologically nontrivial (we called them
``tobogganic") contours ${\cal C}^{(N)} $ which {\em $N-$times
encircle} the strong singularity of our Schr\"{o}dinger equation
in the origin.

In an illustrative presentation of a topologically nontrivial
trajectory ${\cal C}^{(N)}$ let us first recollect that in our
example (\ref{AHO}), Schr\"{o}dinger differential equation will
still be assumed asymptotically harmonic, with its asymptotically
subdominant perturbation $W(ix)$ dominated by the term
$g_{\beta_{max}} \,(ix)^{\beta_{max}}$ with a maximal power
${\beta_{max}}< 2$. This means that equation (\ref{AHO}) will
still possess the two independent asymptotically Gaussian,
harmonic-like solutions,
 \be
 \psi(x) \approx \psi^{(\pm)}(x) =
 e^{\pm x^2/2}\,,\ \ \ \ \ \ \ \ |x| \gg 1\,.
 \label{expo}
 \ee
As long as we may expect a generic, irrational value of $\ell$ we
may treat these solutions of eq. (\ref{AHO}) as multivalued
analytic functions defined on a multi-sheeted Riemann surface with
a branch point at $x=0$. This means that along any ray $x_\theta =
\varrho e^{i\,\theta}$ with the large $\varrho \gg 1$ and at
almost any angle $\theta$ we may re-label solutions (\ref{expo})
as ``physical" [i.e., asymptotically vanishing $\psi^{(phys)}(x)$]
and ``unphysical" [i.e., asymptotically ``exploding"
$\psi^{(unphys)}(x)$]. It is easy to verify that we have,
explicitly,
 \be
  \psi^{(-)}(x) =
  \left \{
  \begin{array}{ll}
  \
  \psi^{(phys)}(x), \ \ \ \ &
  k\pi+\theta \in
  \left (
 -\frac{\pi}{4},\frac{\pi}{4}
  \right ),\\
  \
  \psi^{(un\!phys)}(x), \ \ \ \ &
  k\pi+\theta \in
  \left (
 \frac{\pi}{4},\frac{3\pi}{4}
  \right )
 \ea
 \right ., \ \ \ \ \ \ \ k \in Z\!\!\!Z
 \label{negative}
 \ee
and, {\it vice versa},
 \be
  \psi^{(+)}(x) =
  \left \{
  \begin{array}{ll}
  \
  \psi^{(un\!phys)}(x), \ \ \ \ &
  k\pi+\theta \in
  \left (
 -\frac{\pi}{4},\frac{\pi}{4}
  \right ),\\
  \
  \psi^{(phys)}(x), \ \ \ \ &
  k\pi+\theta \in
  \left (
 \frac{\pi}{4},\frac{3\pi}{4}
  \right )
 \ea
 \right ., \ \ \ \ \ \ \ k \in Z\!\!\!Z\,.
 \ee
Once we restrict our attention to the more usual scenario
(\ref{negative}) we are now prepared to extend the definition
(\ref{BGC}) of the straight-line version ${\cal C}^{(0)} $ of the
``complexified coordinates" $x \in {\cal
D}_{\varepsilon}^{(PTSQM)}$  to the Riemann-surface values of the
``tobogganic trajectories"
 \be
  {\cal D}_{(\varepsilon,N)}^{(PTSQM,\,tobogganic)} =
 \left \{
 x = \varepsilon\,
 \varrho(\varphi,N)\,e^{i\,\varphi} \,  \left | \,\,
 \varphi \in
 \left (
 -(N+1)\pi,\,N\pi^{}
 \right )\,,\ \
  \varepsilon > 0 \right .
 \right \}\,
 \label{BGCN}
 \ee
at any positive integer $N>0$, using a suitable function $
\varrho(\varphi,N)$ in the way discussed more thoroughly in ref.
\cite{tobog}. For the sake of definiteness  we may parametrize
 \ben
  \ \varrho(\varphi,N)=\sqrt{1 + \tan^2 \frac{\varphi+\pi/2}
 {2N+1}}\,.
 \een
This generalizes the Buslaev's and Grecchi's straight line and the
Tanaka's single-sheet curves $\zeta(t)$ (\cite{Tanaka}, with
$N=0$) to the smooth tobogganic spirals ${\cal C}^{(N)} $ at all
$N> 0$ (in their $N=2$ sample in Figure 1 we choose
$\varepsilon=0.05$).

\section{${\cal PT}-$symmetry in the presence of
one or more branch points \label{II.3}}

In the complex plane ${\cal K}$ of $x$ equipped with an
upwards-oriented cut starting at $x=0$ the complex-conjugation
operation ${\cal T}: f(x) \to f^*(x)\equiv \tilde{f}(x^*)$ is
easily applied to the potentials $V(x)$ as well as to the related
wave functions $\psi(x)$. In contrast, the introduction of an
appropriately complexified parity operator ${\cal P}: x \to -x$
seems less straightforward as its action upon the line ${\cal
C}^{(0)}$ proves discontinuous along the negative imaginary
half-axis in ${\cal K}$.

The doublet of the complex parity-like operators ${\cal
P}^{(\pm)}: x \to x \cdot \exp (\pm i\pi)$ might be preferred
since both of them remain continuous. This is achieved at the
expense of having some $x \in {\cal K} ={\cal K}_0 $ mapped out of
the space ${\cal K}$, i.e., by definition, into the neighboring
Riemann sheets ${\cal K}_{\pm 1}$.

In the language of algebra the less immediate invertibility of the
new operators $ {\cal P}^{(\pm)} \neq \left ({\cal P}^{(\pm)}
\right )^{-1}$ makes them even less similar to the standard parity
involution. At the same time their action remains independent of
our (artificially set) cuts so that its definition and/or
visualization becomes facilitated when our (single) cut is
suitably rotated by an angle $\beta$ (${\cal K}\to {\cal
K}_\beta$, cf. a few illustrative pictures in ref. \cite{tobog}).

After we admit a rotation of the cut, the new problem emerges
since also the antilinear action of the operator ${\cal T}$ may
only be well defined  on the whole atlas of the Riemann sheets.
One of the two eligible rotation-type innovations ${\cal
T}^{(\pm)}$ should be considered again, and the same conclusion
must finally be also applied to the product of the operators
${\cal P}^{(\pm)}{\cal T}^{(\pm)}$.

\subsection{Toboggans in potentials with a single spike \label{IV.1}}

In the purely technical sense the explicit (say, numerical)
construction of bound states $\psi^{(N)}(x)$ in the ``$N-$th
quantum toboggan" may be made complicated by the changes of the
related concepts of the parity  and time reversal which become
non-involutive, ${\cal P}^{(\pm)}\neq \left [{\cal
P}^{(\pm)}\right ]^{-1}$ and ${\cal T}^{(\pm)} \neq \left [ {\cal
T}^{(\pm)}\right ]^{-1}$. In this context, our specific choice of
the asymptotically quadratic asymptotics of the potentials of eq.
(\ref{AHO}) and of their manifestly ${\cal PT}-$symmetric form
$V(x)= x^2+\lambda W(ix)$ with rational exponents $\beta$ admits a
certain clarification medited by the following  ${\cal
PT}-$symmetry-preserving change of the coordinates \cite{classif},
 \be
 ix = (iy)^{\tau}, \ \ \ \ \psi(x) = y^{({\tau}-1)/2}\,\Psi(y)\,.
 \label{chov}\,.
  \ee
It may lead to the whole series of the``new", equivalent
re-arrangements of our ``old" Schr\"{o}dinger equation
(\ref{AHO}). Some of them are exceptional as containing solely
even powers of the ``new" complex coordinate,
 \be
 \left [
 -\frac{d^2}{dy^2} + \frac{{\cal L}({\cal L}+1)}{y^2} +
 G_1y^2+G_2y^4 + \ldots + G_{K}y^{2K}
 \right ] \,\Psi (y)= \varepsilon \,\Psi (y)\,.
   \label{trueAHO}
 \ee
Both the Sturm-Liouville bound-state problems (\ref{AHO}) and
(\ref{trueAHO}) may be tobogganic. They are equivalent by
construction. The choice of the constant ${\tau}$ is dictated by
the requirement that for {\em all} the rational exponents $\beta$
which appeared in the original, asymptotically subharmonic
anharmonicity $W(ix)$ at non-zero couplings $g_{(\beta)}$ we get
the {\em even} powers in the new polynomial bound-state problem
(\ref{trueAHO}),
 \ben
 {\beta\cdot{\tau}} = 2m_{(\beta)}, \ \ \ \ \
 m_{(\beta)} = {\rm integer\ for \ any}\ \beta\,.
 \een
We see that the old asymptotically dominant interaction term $x^2$
is replaced by the new maximal-power term
$-{\tau}^2(iy)^{4{\tau}-2}$ (i.e., ${\tau}$ must be integer or
half-integer). The original energy term is tractable as an
interaction term proportional to a constant $x^0$, i.e., we may
set $-E=\lambda\,g_{(0)}x^0$. In this way the constant $E$
acquires the role of one of the coupling constants after the
transformation, $G_{{\tau}-1} = (-1)^{\tau} \lambda{\tau}^2E$. In
the opposite direction, one gets the ``new" energy term from the
old coupling $g_{(\beta_{{\tau}})}$ (which may be, incidentally,
equal to zero) with $\beta_{\tau} = -2+2/{\tau}$. Similarly, the
old and new centrifugal terms must obey the rule
 \ben
 {\tau}^2\left (\ell + \frac{1}{2} \right )^2 = \left ( {\cal L} +
 \frac{1}{2} \right )^2\,.
 \een
Due to the conservation of ${\cal PT}-$symmetry by the change of
variables (\ref{chov}) the ``new" spiked polynomial interaction
representation (\ref{trueAHO}) of our toboggan (\ref{AHO}) (using,
for economy reasons, the minimal parameter ${\tau}$) must have
real coupling constants and will be called ``canonical" in what
follows.

Transformation (\ref{chov}) of coordinates induces the replacement
of the ``old" contour $\left [ x \in {\cal C}^{(N)}\right ]$ by
its ``new" map $\left [ y \in {\cal C'}^{(N')}\right ]$ where, by
construction, $N' \leq N$. In this way the transformation may
lower the winding number. In principle, one may even employ a
``very large" ${\tau}$ in (\ref{chov}) [which need not be
(half)integer in such a case] and achieve a replacement of {\em
any} single-branch-point {\em tobogganic} operator of eq.
(\ref{AHO}) with $N > 0$ by its formally equivalent {\em
non-tobogganic} representation (\ref{trueAHO}) with $N'=0$.

In an alternative presentation we may select a specific ``old"
${\cal PT}-$symmetric tobogganic domain ${\cal D}_{(\varepsilon,
N)}^{(PTSQM,\,tobogganic)}$ such that the equivalence
transformation (\ref{chov}) replaces it by the Buslaev's and
Grecchi's {\em straight line} ${\cal C}^{(0)}$. In the other
words, all the subtleties arising in connection with the
definition of the ${\cal PT}-$symmetry may entirely easily be
solved by their pull-back from the non-tobogganic canonical
Schr\"{o}dinger eq. (\ref{trueAHO}).

We may summarize that the original left-right-reflection-symmetric
interpretation of the ${\cal PT}-$symmetry of any $\psi(x)$
(defined, in ${\cal K}_0$, on the straight line ${\cal C}^{(0)}$
with the geometric center at ${\delta}=\varphi +\pi/2=0$) [in the
notation of eq. (\ref{BGCN})] finds its easy and intuitively
appealing generalization in the ${\delta} \to -{\delta}$ geometric
symmetry of the spirals of the type ${\cal C}^{(N)}$ with respect
to their ``main vertex" at $\varphi = -\pi/2$. For our purposes
each tobogganic spiral ${\cal C}^{(N)}$ may be assigned its
conjugate partner by using just the (arbitrarily selected) action
of one of the eligible rotations ${\cal T}^{(\pm)}$,
 \be
  \left ({\cal C}^{(N)}\right )^\dagger
  ={\cal D}_{(\varepsilon',N)}^{(PTSQM,\,tobogganic)}\,,
  \ \ \ \ \ \
 \varepsilon'=\varepsilon\cdot e^{+ i\pi}\ \ \ {\rm or} \ \ \
 \varepsilon'=\varepsilon\cdot e^{- i\pi}
  \,.
 \label{BGCNcc}
 \ee
The non-tobogganic, standard complex conjugation is re-obtained at
$N=0$.

\subsection{Toboggans in potentials with more spikes \label{IV.2}}

In ref. \cite{can} another hidden source of emergence of a
centrifugal singularity can be spotted in the reconstruction of a
spiked $V_1(x)$ from its {\em regular} supersymmetric partner
${V}_0(x)$. This idea has further been pursued by Sinha and Roy
\cite{ASPR} who started from a regular ${\cal PT}-$symmetric
harmonic oscillator ${V}_0^{(HO)}(x)$ and proposed the
construction of the whole series of potentials $V^{(HO)}_k(x)$
with $k$ different centrifugal-like singularities on the real
line. We may briefly summarize their recipe as specifying the two
supersymmetric partner potentials
 \be
 V_\pm (x) = W^2(x) \pm W'(x)
 \label{susyp}
 \ee
in terms of their shared superpotential $W(x)$ which is, in its
turn, defined by the well known formula \cite{Khare}
 \be
 W(x)=-\frac{\psi'_m(x)}{\psi_m(x)}
 \ee
in terms of an ``input" wave function $\psi_m(x)$. The point is
that {\em any} wave function can be used now, due to the
regularization effect of the ${\cal PT}-$symmetrization
\cite{reguja}. In this sense, the ($m-$plet) of the (simple) nodal
zeros of the $m-$the excited state $\psi_m(x)$ becomes converted
into the simple poles of $W(x)$ (the zeros cannot cancel due to
the Sturm-Liouville oscillation theorem) and into the second-order
poles of at least one of the potentials in (\ref{susyp}) (one
should keep in mind that the poles brought by {\em both} the terms
in (\ref{susyp}) can - and often do - cancel).

It is now easy to imagine that the generic Riemann surface
pertaining to the generic analytic wave function $\psi(x)$ will
have two branch points (say, in $x = \pm 1$) at $m=2$ (etc).
Unfortunately, the corresponding potentials (say,
 \be
 V(x) = x^2+ \frac{G}{(x-1)^2} + \frac{G^*}{(x+1)^2}
 \ee
etc) do not seem to admit closed-form solutions in any generic
case with irrational $G$. Hence, any future analysis of the
related bound states will have to rely upon sophisticated
numerical methods \cite{myinprep}.

An exhaustive analysis of the possibilities of the construction of
the left-right symmetric tobogganic trajectories which would
encircle the {\em pair} of branch points $x=\pm 1$ would be
complicated. One of reasons is that {\em both} the halves of these
trajectories must be permitted to travel freely between the two
singularities before they finally escape in infinity.

The problem of classification of all the possible topologically
nontrivial tobogganic trajectories would become even more
interesting in the case of more than two branch points in the
bound state wavefunctions $\psi(x)$. At the same time, this
problem looks still rather academic on our present level of
knowledge. For this reason, let us now rather turn our attention
to the (as far as we know, equally open) questions connected with
the possibilities of a tobogganic generalization(s) of the usual
textbook {\em scattering} states.

\section{Scattering theory for toboggans \label{III}}


\subsection{The concept of scattering along curved
trajectories \label{III.1}}

Up to now we kept the ${\cal PT}-$symmetric version of Quantum
Mechanics of toboggans specified by the bound-state boundary
conditions at both the ends of the curves ${\cal C}^{(N)}$, i.e.,
in the light and notation of eq.~(\ref{negative}), by the pairs of
the constraints
 \be
 \psi \left ( \varrho \cdot e^{i\,\theta} \right ) = 0,
 \ \ \ \ \ \ \ \varrho \gg 1
  \label{bc}
 \ee
where
 \ben
 \theta + k_{sub}\,\pi \in
  \left (
 -\frac{\pi}{4},\frac{\pi}{4}
  \right )\,.
  \een
Two different subscripts $_{sub} = _{in}$ and $_{sub} = _{out}$
were considered, marking the angle of the ``initial" and ``final"
Stokes-line ray, respectively. This means that we have to select
$k_{out}=0\ {\rm and}\ k_{in}=1\ {\rm at}\ N=0$, $k_{out}=-1\ {\rm
and}\ k_{in}=2\ {\rm at}\ N=1$, $k_{out}=-2\ {\rm and}\ k_{in}=3\
{\rm at}\ N=2$, etc.

Now, let us search further inspiration in certain reflectionless
(or, if you wish, standing-wave) real-line Hermitian models as
studied in letter \cite{Berry}. We will contemplate their
generalization to our present ${\cal PT}-$symmetric tobogganic
context. In fact, the generalization is not difficult as it is
sufficient to combine our present differential Schr\"{o}dinger
equation
 \be
 H_{({\cal PT})}\,\psi(x)=E\,\psi(x)
 \ee
with the appropriately generalized {\em scattering} boundary
conditions. This means that we require that the incoming beam
moves along a specific (often called ``anti-Stokes") line,
 \be
 \psi\left ( \varrho \cdot e^{i\,\theta_{in}} \right ) =
 \psi_{(i)}(x) + B\,\psi_{(r)}(x),
 \ \ \ \ \ \varrho \gg 1, \ \ \ \ \theta_{in}={\rm fixed}.
 \ee
This scattering-like wave function is composed of the normalized
incident wave $ \psi_{(i)}(x) \approx e^{i\varrho^2/2}$ in
superposition with the reflected $ \psi_{(r)}(x) \approx
e^{-i\varrho^2/2}$. Similarly we select the outcoming beam
 \be
 \psi\left ( \varrho \cdot e^{i\,\theta_{out}} \right ) =
 (1+ F)\,\psi_{(t)}(x),
 \ \ \ \ \ \varrho \gg 1, \ \ \ \ \theta_{out}={\rm fixed}
 \ee
which contains just the transmitted wave $ \psi_{(t)}(x) \approx
e^{i\varrho^2/2}$. This notation generalizes the standard textbook
scattering on the real line and could be also reformulated in the
language of the transfer and/or monodromy matrices \cite{Arnold}
in an extension to the case where our complex path of coordinates
is different form the standard straight real line. Our present
notation $B$ and $F$ reminds the reader of the coefficients
called, respectively, the ``backward scattering" and ``forward
scattering" amplitudes in physics.

For our present, asymptotically $x^2-$dominated potentials
(\ref{spiked}) with $\beta_{max}<2$ we can specify the ``in" and
``out" $\varrho\to \infty$ asymptotics of the respective
lower-edge and upper-edge ${\cal PT}-$symmetric scattering curves
${\cal A}^{(N)}_{(L)}$ and ${\cal A}^{(N)}_{(U)}$ in closed form,
 \be
 \begin{array}{ll}
 {\cal A}^{(N)}_{(L)} \to \varrho\,e^{i\,\theta_{in}},
 \ \ \ \ \ & \theta_{in} = -(N+3/4)\,\pi,\\
 {\cal A}^{(N)}_{(L)} \to \varrho\,e^{i\,\theta_{out}},
 \ \ \ \ \ & \theta_{out} = (N-1/4)\,\pi,\\
 {\cal A}^{(N)}_{(U)} \to \varrho\,e^{i\,\theta_{in}},
 \ \ \ \ \ & \theta_{in} = -(N+5/4)\,\pi,\\
 {\cal A}^{(N)}_{(U)} \to \varrho\,e^{i\,\theta_{out}},
 \ \ \ \ \ & \theta_{out} = (N+1/4)\,\pi\,.
 \ea
 \label{BGCNsca}
 \ee
These choices preserve the ${\cal PT}-$symmetry of the new
anti-Stokes tobogganic scattering contours ${\cal A}^{(N)}$
obtained as the edge-of-the wedge boundaries, i.e., the respective
upper {\em or} lower limiting extremes ${\cal A}^{(N)}_{(U)}$ or $
{\cal A}_{(L)}^{(N)}$ of the preceding deformable bound-state
contours ${\cal C}^{(N)}$ at a given $N$.

An extension of this construction to the potentials asymptotically
dominated by the other powers of the coordinate may be based on
the use of formula (\ref{chov}) and is left to the reader.

\subsection{Illustration: Tobogganic scattering by the
harmonic-oscillator
well
 \label{III.2}}

For illustration let us now omit the perturbation $W(ix)$ from
eq.~(\ref{AHO}) and discuss the solution of the resulting
simplified Schr\"{o}dinger differential equation
 \be
 \left [
 -\frac{d^2}{dx^2} + \frac{\alpha^2-1/4}{x^2} + x^2
   \right ] \,\psi (x)= E \,\psi (x), \ \ \ \ \ \ \
   \alpha = \ell+\frac{1}{2},
   \label{HOsca}
 \ee
say, along the path ${\cal A}_{(L)}^{(0)}$, i.e., in the first
nontrivial scattering regime. In the first step let us notice the
$x \leftrightarrow -x$ and $\alpha \leftrightarrow -\alpha$
symmetries of eq.~(\ref{HOsca}) and recollect that on the
asymptotic parts of the selected scattering trajectory ${\cal
A}_{(L)}^{(0)}$ we may set $x^2 = -i r$ with the real $r\ll -1$
along the asymptotic part of the ``in" branch and with $r\gg +1$
for the ``out" branch, respectively.

In the second step we shall set $E=2\mu$ and restrict our
attention to the generic case, ignoring, in a way paralleling the
bound-state construction \cite{ptho}, the exceptional integer
values of $\alpha$. This allows us to construct the general
analytic solution $\psi(x)$ of our ordinary differential
Schr\"{o}dinger equation (\ref{HOsca}) of the second order, {\em
in the scattering regime}, as a superposition of the expression
proportional to a confluent hypergeometric function,
 \be
 \chi_{(\alpha)}(r)=r^{\frac{1}{4}+\frac{\alpha}{2}}
 \,e^{ir/2}\,_1F_1\,\left (
 \frac{\alpha+1-\mu}{2}, \alpha+1; -ir
 \right )
 \label{solar}
 \ee
with its linearly independent partner $ \chi_{(-\alpha)}(r)$.

In the third step we may employ the well known $|r| \gg 1$
estimate for hypergeometric functions \cite{Fluegge} and get the
compact final asymptotic formula
 \be
 r^{\frac{1}{4}+\frac{\alpha}{2}}\chi_{(\alpha)}(r) \approx
  \,e^{ir/2}\,
 \frac{r^{\mu/2}\,\exp \left [
 -i\,{\pi}\,
 \left (
 \alpha+1
 \right )/4
 \right ]}{
 \Gamma
 \left [
 (\alpha+1+\mu)/2
 \right ]
 }
 +e^{-ir/2}\,
 \frac{r^{-\mu/2}\,\exp \left [
 +i\,{\pi}\,
 \left (
 \alpha+1
 \right )/4
 \right ]}{
 \Gamma
 \left [
 (\alpha+1-\mu)/2
 \right ]
 }\,.
 \label{solarbe}
 \ee
Its inspection reveals that at the generic $\alpha>0$ and $\mu =
E/2>0$ the dominant asymptotic $|x| = |\sqrt(r)| \gg 1$ behaviour
of the wavefunctions is ``rigid",
 \be
 \psi_{in,out}(x) \approx
 r^{-{1}/{4}+(\alpha+\mu)/2}
 \,e^{ir/2}\,
 \frac{\exp \left [
 -i\,{\pi}\,
 \left (
 -\alpha+1
 \right )/4
 \right ]}{
 \Gamma
 \left [
 (-\alpha+1+\mu)/2
 \right ]
 }+ corrections\,.
 \label{domin}
 \ee
We may conclude that the scattering properties of our model remain
trivial unless the energies acquire specific values at which the
dominant term (\ref{domin}) would vanish. This observation is not
surprising. Indeed, even the ``scattering" boundary conditions may
lead to the quantization of the spectrum of energies in the way
illustrated in ref. \cite{Berry}. Incidentally, the authors of the
latter reference preserved the traditional, non-tobogganic
real-line contour ${\cal A}^{(0)} \equiv I\!\!R$ in their model
``fine-tuned" to the specific asymptotics of their interaction
$V(x) \sim -x^4$. From our present point of view their particular
choice of the contour is in fact not unique. Its ambiguity may
again be interpreted as a generic feature of the analytic models.

In a different perspective our result~(\ref{domin}) resembles the
standard scattering in the Coulomb field \cite{Fluegge} where the
phase-shift of the oscillations of the scattered wave remained
coordinate-dependent due to the not sufficiently rapid asymptotic
decrease of the potential. In the other words, the standard
textbook  Coulombic scattered waves $\psi_{out}^{(Coul)} (r)$ are
``distorted" by a power-law factor as well,
 \ben
    \sin (\kappa r + const) \to \sin (\kappa r + const\cdot \log r
 + const)\,.
 \een
In our eqs.~(\ref{solarbe}) and (\ref{domin}), {\em conceptually
the same} power-law distortion of the ``free waves" $\psi_{in,out}
(r)$ occurs.

For this reason a suitably modified definition of the scattering
matrix ${\cal S}$ (measuring the change of the ratio of two
independent solutions due to the scattering) would be needed.
{Mathematically}, the present construction of the scattering
states for the nontrivial power-law potentials is clear and almost
obvious since our curved contours (\ref{BGCNsca}) represent a very
natural generalization of the usual real line of $x$. At the same
time, a real application/applicability of this construction
represents a {\rm truly} open problem. One must keep in mind that
the coordinates $x$ themselves lie on the  {\em complex} contours
(\ref{BGCNsca}) and cannot be understood as directly observable,
therefore. Presumably, for this reason (and in full analogy with
the parallel ${\cal PT}-$symmetric bound state problems), the
correct {physical} interpretation of our unusual scattering states
generated by the asymptotically nontrivial power-law potentials
would again require the construction and specification of an
appropriate non-local operator of the physical metric $\Theta$
\cite{mujrev}.

\section{Summary and outlook \label{IV}}

One of the most important innovations brought by the ${\cal
PT}-$symmetric Quantum Mechanics is that it decouples its models
and observables (say, quantum Hamiltonians $H$) from an immediate
correspondence to their classical analogues. In particular, once
we allow that the coordinates $x$ become complex (i.e., the
``position" of a ``particle" ceases to be observable), the
experimental observability of these quantities is, naturally,
lost. Their role remains formal and purely auxiliary. Suitable
functions $f(x)$ of these quantities may still be treated as
elements of the Hilbert (or Krein \cite{Tanaka}) space in an
``easily tractable" representation but the values $x$ of their
argument cannot be understood as eigenvalues of an operator of
``position" anymore. In physical context this loss of
observability for some ``obvious operators of coordinates" finds
its closest analogue in relativistic quantum mechanics where one
encounters a contradictory and unobservable ``Zitterbewegung" of
the Dirac particles \cite{Messiah}, etc.

In the mathematical setting the best developed rigorous analysis
(which may be exemplified by the recent paper by Trinh
\cite{Trinh} and references listed therein) exists for polynomial
potentials. Using an explicit reference to the underlying theory
of the corresponding linear differential equations of second order
in complex plane \cite{Sibuya} one may formulate a number of
theorems. Unfortunately, the bulk of the available mathematical
theory has not been extended to an appropriate incorporation of
singular terms in $V(x)$ nor to the problem of scattering.

In such a setting our present paper studied the formal
Hamiltonians of quantum toboggans which do not seem to have any
obvious source or inspiration in Classical Mechanics. In a way
characteristic for modern physics, the real use of their formal
development may only be expected to come {\it a posteriori}. In
this spirit our present paper filled the gap since without an
explicit construction of scattering states, the concept of quantum
toboggans considered just as a bound-state problem along
topologically nontrivial complex trajectories seemed incomplete.

\section*{Acknowledgement}

The work was supported by the Institutional Research Plan
AV0Z10480505 and by the M\v{S}MT ``Doppler Institute" project Nr.
LC06002.

\section*{Figure captions}

\subsection*{Figure 1. Complex trajectory of the toboggan ${\cal C}^{(N)}$
at $N=2$ }

\end{document}